\documentclass[aps,pra,superscriptaddress,twocolumn,longbibliography]{revtex4-2}

\usepackage[T1]{fontenc}
\usepackage[utf8]{inputenc}
\usepackage{url}
\usepackage{amsmath}
\usepackage{mathabx}
\usepackage{mathtools}
\usepackage{bbold}
\usepackage{mathrsfs}
\DeclareFontFamily{U}{dutchcal}{\skewchar\font=45 }
\DeclareFontShape{U}{dutchcal}{m}{n}{<-> s*[1.0] dutchcal-r}{}
\DeclareFontShape{U}{dutchcal}{b}{n}{<-> s*[1.0] dutchcal-b}{}
\DeclareMathAlphabet{\mathlcal}{U}{dutchcal}{m}{n}

\usepackage{graphicx}
\usepackage[pdfstartview=FitH,colorlinks=true,citecolor=blue,linkcolor=blue,urlcolor=blue]{hyperref}
\usepackage{float}
\usepackage{mathptmx}
\usepackage{tikz}
\usepackage[vcentermath]{youngtab}

\begin{document}

\author{Pablo Capuzzi}
\affiliation {Universidad de Buenos Aires, Facultad de Ciencias Exactas y Naturales,
Departamento de Física, Pabell\'on 1, Ciudad Universitaria, 1428 Buenos Aires, Argentina}
\affiliation{CONICET - Universidad de Buenos Aires, Instituto de Física de Buenos
Aires (IFIBA). Buenos Aires, Argentina}
\author{Zehra Akdeniz}
\affiliation{Faculty of Engineering, P\^{\i}r\^{\i} Reis University, 34940 Tuzla, Istanbul, Turkey}
\author{Patrizia Vignolo}
\affiliation{Universit\'e C\^ote d’Azur, CNRS, Institut de Physique de Nice, 06200 Nice, France}
\affiliation{Institut Universitaire de France}

\title{Shortcut to spin dynamics in quantum mixtures}
\begin{abstract}
  The study of the long-time dynamics of quantum systems can be a real
  challenge, especially in systems like ultracold gases, where the
  required timescales may be longer than the lifetime of the system
  itself. In this work, we show that it is possible to access the
  long-time spin dynamics of a strongly repulsive atomic gas mixture in
  shorter times. The shortcut-to-dynamics protocol that we propose
  does not modify the fate of the observables, but effectively jumps
  ahead in time without changing the system's inherent evolution. Just
  like the next-chapter button in a movie player that allows to quickly reach
the part of the movie one wants to watch, it is a leap into the future.
\end{abstract}

\maketitle
Observing and studying quantum phenomena that occur over long
timescales can be a daunting task.  For example, it can be extremely
difficult to distinguish between a very slow thermalization dynamics
and a situation where thermalization does not occur
\cite{Luitz2015,Alet2018}, since this requires observing dynamics over
a very long time, time in which the quantum system could be deeply
perturbed by the coupling with the external environment.  Another
phenomenon that is interesting to study over a long time, and which
can suffer from the decoherence induced by the interaction with the
environment, is the coherence spreading. This observable may give
access, for instance, to new forms of quantum information transfer
\cite{Quiao2020}.

\begin{figure}
    \centering
    \includegraphics[width=0.7\linewidth]{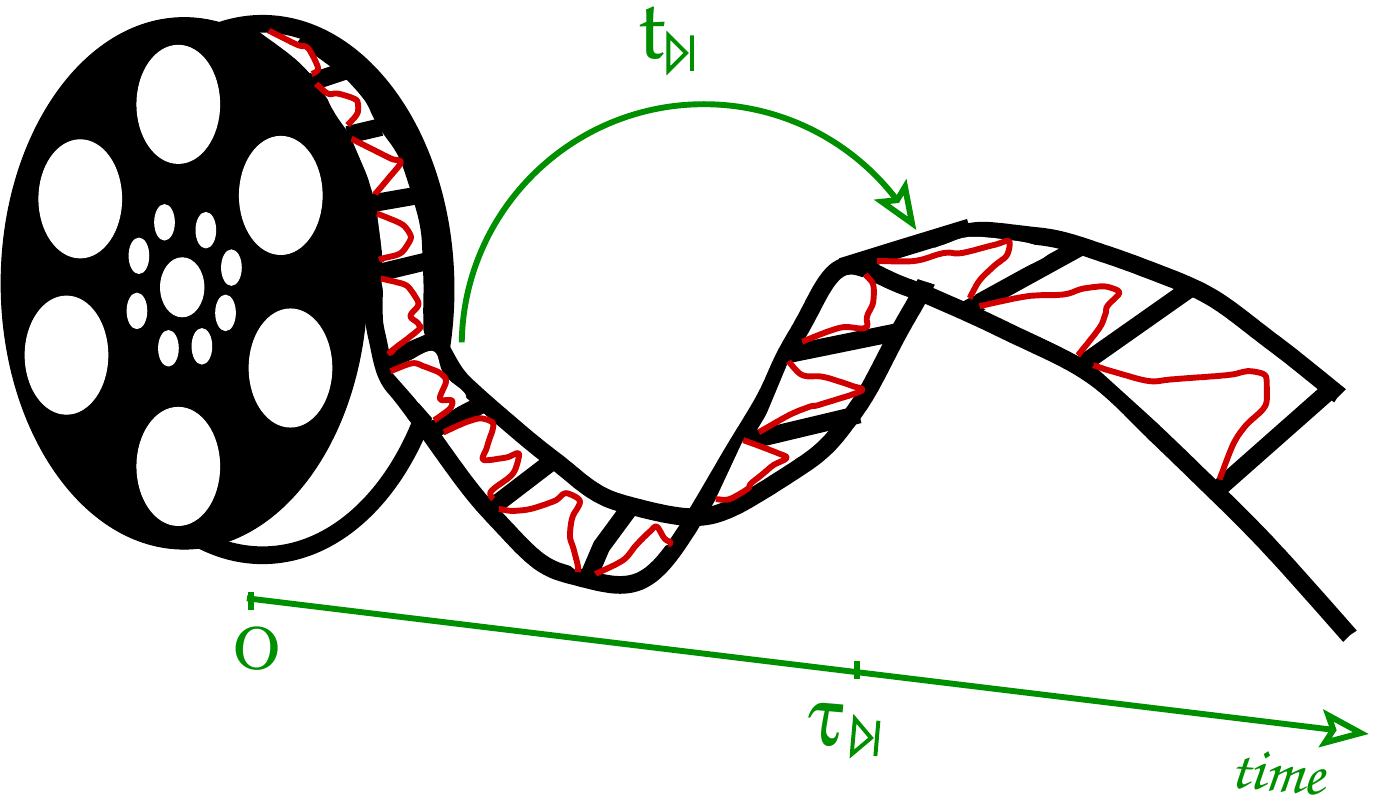}
    \caption{Schematic representation of the 
       shortcut-to-dynamics protocol proposed in this
      work: a time leap $\tau_{\smalltriangleright\scalebox{.5}{|}}$
      realized in a time $t_{\smalltriangleright\scalebox{.5}{|}}$,
      with
      $t_{\smalltriangleright\scalebox{.5}{|}}\ll\tau_{\smalltriangleright\scalebox{.5}{|}}$.}
    \label{idea}
\end{figure}

When experimental timescales become a limiting factor, the question
arises if it is possible to devise a shortcut protocol that fast
forwards the dynamics for a given time, thereby reducing the time
needed to observe a physical phenomenon of interest. The concept of
driving the dynamics has been previously introduced (see for instance
\cite{Masuda2008,Masuda2009,Torrontegui2012,Torrontegui2013,Martinez2016,Patra_2017,GueryOdelin2019,Bernardo2020,Plata2021,GueryOdelin2023}
and references therein), but as a strategy to speed up an adiabatic
evolution to attain a target state or realize a given protocol in a
shorter time
\cite{Chen2010,Schaff2010,Schaff2011,Delcampo2011a,Delcampo2012a,Delcampo2013a,Deng2018,Diao_2018}.
Here the idea is different: fast-forwarding the dynamics to access the
system's dynamics starting from a time
$\tau_{\smalltriangleright\scalebox{.5}{|}}$ after a time
$t_{\smalltriangleright\scalebox{.5}{|}}\ll
\tau_{\smalltriangleright\scalebox{.5}{|}}$. The goal is not to target
a specific quantum state, neither to speed up a specific protocol but
rather to jump ahead to the time when we want to start observing the
dynamics, just as when we use the next-chapter button to watch the
part of the movie that interests us, without needing to know anything
of the movie in advance (see Fig. \ref{idea}).

Our protocol applies to one-dimensional (1D) strongly repulsive
quantum mixtures \cite{Zurn2012}.  Such systems can be mapped onto
Heisenberg spin chains
\cite{deuretzbacher_quantum_2014,Volosniev2016,Aupetit2022} and admit
an exact solution in the limit of infinite interactions
\cite{volosniev_strongly_2014}, so that they are a paradigm for the
study of the long-time dynamics in the vicinity of integrability
\cite{pecci2022,Musolino2024}.  They could also be a good candidate
for the study of the interplay between interactions and disorder
\cite{Capuzzi2024}. In strongly repulsive 1D mixtures, the charge and
the spin degrees of freedom separate \cite{arute2020}.  More
precisely, the dynamics of the different spin components is governed,
in addition to the strength of the interaction and the statistics, by
the total density, whose evolution, however, does not depend on the
inter-component dynamics
\cite{deuretzbacher_exact_2008,volosniev_strongly_2014,deuretzbacher_quantum_2014,pecci2022,Capuzzi2024,Musolino2024}.

In this letter, we show that this separation allows to fast-forward
the spin-components dynamics up to a time
$\tau_{\smalltriangleright\scalebox{.5}{|}}$ by first compressing and
then decompressing the total density during a much shorter time
$t_{\smalltriangleright\scalebox{.5}{|}}$,
$\tau_{\smalltriangleright\scalebox{.5}{|}}$ being a function of
$t_{\smalltriangleright\scalebox{.5}{|}}$,
$\tau_{\smalltriangleright\scalebox{.5}{|}}=\tau(t_{\smalltriangleright\scalebox{.5}{|}})$. If
the compression-decompression cycle is suitably performed such that
the total density recovers its initial size at
$t_{\smalltriangleright\scalebox{.5}{|}}$ without further exciting the
system, then for $t=t_{\smalltriangleright\scalebox{.5}{|}}+t'$ the
system continues the time evolution such as it was at time
$\tau_{\smalltriangleright\scalebox{.5}{|}}+t'$: we get a shortcut to
dynamics for the spin components.

{\it Model -} We consider a fermionic or bosonic spin mixture with
$\kappa$ components, trapped in an 1D time-dependent harmonic trap of
frequency $\omega(t)$.  The mixture obeys the Hamiltonian
\begin{equation}
\begin{split}
\hat{H}&=\sum^\kappa_{\sigma=1}\sum_{i=1}^{N_\sigma}
\left[-\frac{\hbar^2}{2m}\frac{\partial^2}{\partial x_{i,\sigma}^2}+\dfrac{1}{2}m \omega(t)^2 x_{i,\sigma}^2\right]+
\sum^\kappa_{\sigma=1}g_{\sigma\sigma}\\
&\sum_{i=1}^{N_\sigma} \sum_{j>i}^{N_{\sigma}} \delta(x_{i,\sigma}-x_{j,\sigma})+ \frac 1 2
\sum^\kappa_{\sigma\neq\sigma'=1}g_{\sigma\sigma'} \sum_{i=1}^{N_\sigma}\sum_{j=1}^{N_{\sigma'}} \delta(x_{i,\sigma}-x_{j,\sigma'}),\\
\end{split}
\label{ham}
\end{equation}
where $\kappa$ is the number of spin species, $N_{\sigma}$ the number
of particles with spin $\sigma$, $g_{\sigma\sigma'}$ and
$g_{\sigma\sigma}$ are the inter- and intra-species interaction
strengths. The latter concerns only identical bosons, since for
identical fermions {\it s}-wave contact interactions are forbidden.
The regime we are interested in is the strongly repulsive one, where
$g_{\sigma\sigma}m/(\hbar^2n(x=0))\gg 1$ and
$g_{\sigma\sigma'}m/(\hbar^2n(x=0))\gg 1$, for any $\sigma,\sigma'$,
where $n(x=0)$ is the density at the trap center.

The proposed protocol applies to any mixture that obeys the
Hamiltonian (\ref{ham}) in such a regime, one of the key points being
the spin-density separation that occurs for strongly repulsive
interactions. Nevertheless, in the following, we will focus on the
case of an SU(2) balanced fermionic mixture ($\kappa=2$,
$g_{\uparrow\uparrow}=g_{\downarrow\downarrow}$,
$g_{\uparrow\downarrow}=g$ and $N_\uparrow=N_\downarrow=N/2$).  The
system evolution we want to measure at long times is that of the two
spin-components in a static harmonic confinement \cite{pecci2022}. The
shortcut to dynamics idea is to drive the proper time of the spin
components by modifying the harmonic trapping during a short time. For
that matter, we will construct the many-body wavefunction for such
time-dependent Hamiltonian starting from the many-body wavefunction of
the time-independent one. This will allow us to engineer our
shortcut-to-dynamics protocol.

{\it The evolution for a time-independent Hamiltonian -} As a first
step, we consider the evolution of the system under time-independent
harmonic trapping, $\omega(t)=\omega_0$, $\forall t$, in the absence
of charge excitations.  In the strongly repulsive regime, the
many-body wavefunction, $\Psi(\vec{X}, \vec{\sigma}, t)$ with
$\vec{X} = \{x_1, x_2, \dots, x_N \}$ and
$\vec{\sigma} = \{\sigma_1, \sigma_2, \dots, \sigma_N \}$, vanishes
whenever $x_i=x_j$.  It can be written as
follows~\cite{deuretzbacher_exact_2008,volosniev_strongly_2014}
 \begin{equation}
  \Psi(\vec{X}, \vec{\sigma}, t)= \sum_{P\in S_N}   \alpha_P(t) \theta_P(\vec{X})\Xi_A(\vec{X}),
  \label{eq:Psi_MB}
  \end{equation}
  where the $\vec{\sigma}$-dependence is hidden in the spin
  coefficients $\alpha_P(t)$, with $P$ permutation of $N$ elements in
  $S_N$. $\theta_P(\vec{X})$ is the generalized Heaviside function,
  which is equal to 1 in the coordinate sector
  $x_{P(1)}<\dots<x_{P(N)}$ and 0 elsewhere, and
  $\Xi_A=(1/\sqrt{N!}) \det [\chi_j(x_k)]$ is the Slater determinant
  built from the orbitals $\chi_j(x)$ of the single-particle
  Hamiltonian for the harmonic potential,
  $H_0\chi_j(x)=\varepsilon_j \chi_j(x)$, with
\begin{equation}
    H_0=-\frac{\hbar^2}{2m}\frac{\partial^2}{\partial x^ 2}+\dfrac{1}{2}m \omega_0^2 x^2.
\end{equation}
Since there are no charge excitations, $\Xi_A$ is time independent,
and the many-body evolution is dictated by $\alpha_P$.

The time evolution of the $\alpha_P(t)$'s coefficients is determined
by the effective spin-chain Hamiltonian
~\cite{deuretzbacher_quantum_2014}
\begin{equation}
 H_S^0=\sum_{j=1}^{N-1}\left( -J_j^0\hat{1}+J_j^0\hat P_{j,j+1}\right)
\label{spin_chain}
\end{equation}
where $\hat P_{j,j'}$ is the permutation operator.  The hopping
amplitudes $J_i^0$ in Eq.~(\ref{spin_chain}) can be written as
\begin{equation}
    J_i^0 = \frac{N!}{g} \int_{-\infty}^{\infty} dX ~\delta(x_i - x_{i+1})  \theta_{\text{id}}(\vec{X}) \Bigl\lvert \frac{\partial \Xi_A}{\partial x_i}\Big\lvert^2,
\label{eq:Ji}
\end{equation}
where $\Theta_{\text{id}}(\vec{X})$ is the generalized Heaviside
function for the identity permutation.  Let $\varepsilon_{S,j}$ and
$\alpha_P^j$ be the eigenvalues and the corresponding eigenvectors of
$H_s^0$. If we write $\alpha_P(t=0)=\sum_j\beta_j\alpha_P^j$, then the
evolution of $\alpha_P$ is given by
\begin{equation}
    \alpha_P(t)=\sum_j \beta_j \alpha_P^j e^{-i\varepsilon_{S,j}t/\hbar}.
\end{equation}

Such a system has a time-independent total density
\begin{equation}
    n_0(x)=\int dx_2\,dx_3\dots dx_N|\Xi_A(x,x_2,\dots x_N)|^2
\end{equation}
and spin-component densities
\begin{equation}
    n_{\sigma,0}(x, t) =\sum_{i=1}^N\left(\sum_P|\alpha_p(t)|^2\delta_{\sigma,\sigma_i}\rho_0^{(i)}(x)\right)
\end{equation}
where the density of the $i-$th fermion reads
\begin{equation}
    \rho_0^{(i)}(x)=N!\int_{\mathcal{I}_i} \prod_{j\neq i}|\Xi_A(x_1,\dots x_{i-1},x,x_{i+1},\dots,x_N )|^2,
\end{equation}
 $\mathcal{I}_i$ being the integration interval constrained by $x_1<x_2<\dots<x_{i-1}<x<x_{i+1}<\dots<x_N$.

 If the initial spin state is not an eigenstate of the $H_S^0$, even
 if the total density is constant in time, there is a relative motion
 between the spin components
 \cite{wei2021quantum,pecci2022,Musolino2024}, so that the density
 magnetization
 $m_0(x, t) = n_{\uparrow,0}(x, t)-n_{\downarrow,0}(x, t)$ is not
 constant in time.  Following the time-evolution of its center of mass
\begin{equation}
    d_0(t)=\frac{1}{N}\int dx\,x\,m_0(x,t).
\end{equation}
one can have access to interesting physical phenomena related to the
motion of the spin components such as superdiffusion at short time
\cite{wei2021quantum,pecci2022}, universal oscillations at
intermediate time \cite{pecci2022}, and eventually thermalization at
long times \cite{pecci2022,Capuzzi2024}.  In the strongly interacting
regime, the timescale for the spin’s motion is $\hbar/J_i^0$, namely,
it scales with the particle's interaction strength so that it can be
very large with respect to the harmonic oscillator period.  The
observation of the spin dynamics at long times can therefore become
difficult to access due to the finite lifetime of the trapped system.
Our goal will be to have access to the spin dynamics for
$t>\tau_{\smalltriangleright\scalebox{.5}{|}}$ after a
time %$t_{\smalltriangleright\scalebox{.5}{|}}$,
with
$t_{\smalltriangleright\scalebox{.5}{|}}\ll
\tau_{\smalltriangleright\scalebox{.5}{|}}$. In particular, we shall
focus our analysis on the evolution of the spin-component densities
$n_{\sigma,0}(x, t)$ and the magnetization center of mass $d_0(t)$.

{\it The evolution for the time-dependent Hamiltonian -} When the trap
frequency depends on time, the evolution of the many-body wavefunction
admits a scaling solution that keeps the same form as the one for the
time-independent problem
\begin{equation}
  \Psi(\vec{X}, \vec{\sigma}, t)= \sum_{P\in S_N} a_P(t) \theta_P(\vec{X})\Psi_A(\vec{X},t),
  \label{eq:Psi_MB-bis}
  \end{equation}
Here
\begin{equation}
 \Psi_A(\vec X,t)=\dfrac{1}{b^{N/2}}\Xi_A(\vec{Y})\,e^{\displaystyle-\frac{i}{\hbar}\int_0^t \frac{dt'}{b^2}\sum_{j=0}^{N-1}\varepsilon_j}\times e^{\displaystyle i\frac{m}{\hbar}\frac{\dot b}{2b}\sum_{k=1}^N x_k^2}  
\end{equation}
is the Slater determinant of the space-and-time rescaled
time-dependent one-particle orbitals
\cite{Minguzzi2005,Gritsev2010,Chen2010,delCampo2011,Schaff2011,GueryOdelin2019}
\begin{equation}
\phi_j(x_k,t)=\dfrac{1}{\sqrt{b}}\chi_j(y_k)e^{\displaystyle-\frac{i}{\hbar}\varepsilon_j\int_0^t \frac{dt'}{b^2}}\,e^{\displaystyle i\frac{m}{\hbar}\frac{\dot b}{2b}x_k^2}
\end{equation} 
where $y_k=x_k/b$ and the scaling function $b(t)$ verifies the Ermakov equation
    \begin{equation}
        \ddot b+\omega^2(t)b=\dfrac{\omega_0^2}{b^3}.
        \label{Ermakov}
    \end{equation}
%    Remark that the natural time scaling for the orbitals, and thus for the spatial part
%    of the many-body wavefunction is $\tau=\int_0^t dt'1/(b(t')^2$.
    The Slater determinant scaling allows us to obtain the scaling
    properties of the hopping amplitudes $J_i(t)$ governing the spin
    dynamics \cite{Volosniev2016}:
\begin{equation}
\begin{split}
      J_i(t) &=  \frac{N!}{g}  \int_{-\infty}^{\infty} dX ~\delta(x_i - x_{i+1})  \theta_{\text{id}}(X) \Bigl\lvert \frac{\partial \Psi_A}{\partial x_i}\Big\lvert^2\\
      &=\frac{N!}{g}  \int_{-\infty}^{\infty} dX ~\delta(x_i - x_{i+1})  \theta_{\text{id}}(X) \frac{1}{b^{N+2}}\Bigl\lvert \frac{\partial \Xi_A}{\partial y_i}\Big\lvert^2\\     
      &=\frac{N!}{g b^3} \int_{-\infty}^{\infty} dY ~\delta(y_i - y_{i+1})  \theta_{\text{id}}(Y) \Bigl\lvert \frac{\partial \Xi_A}{\partial y_i}\Big\lvert^2\\
      &=\dfrac{J_i^0}{b^3}.
      \end{split}
\end{equation}
This implies that the sector amplitudes $a_P$ can be expressed as a
function of the ones for the time-independent Hamiltonian $\alpha_P$
by rescaling the time. Namely, $a_P(t)=\alpha_P(\tau)$, with
$\tau=\int_0^t dt'/b(t')^3$. This means that by manipulating the
scaling function $b(t)$ through a suitable protocol for $\omega(t)$,
one can observe the spin dynamics of the time-independent Hamiltonian
at different times $\tau$.
%Remark that
%\begin{equation}
 %   \dfrac{\partial\tau}{\partial\tau}=\dfrac{1}{b},
%\end{equation}
%namely the time evolution of the spatial part and of the spin-part are not governed by the same
%scaling laws.

For the time-dependent problem, the total density reads
\begin{equation}
\begin{split}
    n_0(x,t)&=\int dx_2\,dx_3\dots dx_N|\Psi_A(x,x_2,\dots x_N)|^2\\
    &=\dfrac{1}{b}n_0(x/b)
    \end{split}
\end{equation}
and the density of the $i-$th fermion scales
as $\rho^{(i)}(x,t)=1/b\,\rho_0^{(i)}(x/b)$.
Thus, the spin component densities can be written as
\begin{equation}
\begin{split}
    n_{\sigma}(x, t) &=\sum_{i=1}^N\left(\sum_P|a_p(t)|^2\delta_{\sigma,\sigma_i}\rho^{(i)}(x,t)\right)\\
    &=\frac{1}{b}\sum_{i=1}^N\left(\sum_P|\alpha_p(\tau(t))|^2\delta_{\sigma,\sigma_i}\rho_0^{(i)}(x/b)\right)\\
    &=\dfrac{1}{b}n_{\sigma,0}(x/b,\tau(t)).
    \label{spindensscaled}
    \end{split}
\end{equation}
Analogously, the magnetization can be written as
\begin{equation}
    m(x,t)=\frac{1}{b}m_0(x/b,\tau(t))
\end{equation}
and its center of mass
\begin{equation}
    \begin{split}
        d(t)&=\frac{1}{N}\int dx\,x\,m(x,t)\\
        &=b(t) d_0(\tau(t)).
    \end{split}
    \label{ddt}
\end{equation}

Remark that if the initial state is an eigenstate of $H_S^0$, it will
be an eigenstate of $H_S^0/b^3$ too. In such a case
$\alpha_p(\tau(t))=\alpha_p^j$, so that the magnetization density
follows the same scaling law as the total density, namely
$m(x,t)=m_0(x/b)/b$.

{\it Shortcut-to-dynamics protocol for the spin dynamics -} We seek a
scaling transformation $b(t)$, and thus a time-dependent trap
frequency $\omega(t)$ that can be experimentally realized and that
allows to reach our shortcut-to-dynamics goal.  We choose the strategy
to fast compress and decompress the total density using a shortcut
protocol \cite{Chen2010,Schaff2010,Schaff2011}, so that the initial
density and the final density for
$t\ge t_{\smalltriangleright\scalebox{.5}{|}}$ are identical, ie,
$b(0)=1$ and $b(t\ge t_{\smalltriangleright\scalebox{.5}{|}})=1$, but
the elapsed time for the spin dynamics is significantly longer than
the time duration of the compression-decompression cycle.  The
shortcut protocol implies that $\dot{b}(0)=0$,
$\dot{b}(t_{\smalltriangleright\scalebox{.5}{|}})=0$, $\ddot{b}(0)=0$,
$\ddot{b}(t_{\smalltriangleright\scalebox{.5}{|}})=0$, while the
fast-forward condition is fulfilled if
$b(0<t<t_{\smalltriangleright\scalebox{.5}{|}})<1$.

As a specific example, we choose the compression and the decompression
times to be equal, by imposing
$\dot{b}(t_{\smalltriangleright\scalebox{.5}{|}}/2)=0$.  Proposing a
lower order polynomial solution to the Ermakov equation, we can write
a solution for the scaling parameter $b$ as
\begin{equation}
b(t)=1-|A|\left(-({t}/{t_{\smalltriangleright\scalebox{.5}{|}}})^6+3(t/t_{\smalltriangleright\scalebox{.5}{|}})^5-3(t/t_{\smalltriangleright\scalebox{.5}{|}})^4+(t/t_{\smalltriangleright\scalebox{.5}{|}})^3\right),
\end{equation}
which, in turn, determines the trapping strength through
Eq. (\ref{Ermakov}).  The amplitude $|A|$ has to be chosen so that
$\omega^2(t)>0$ to ensure the system remains confined during the
shortcut protocol.

\begin{figure}
    \centering
    \includegraphics[width=0.8\linewidth]{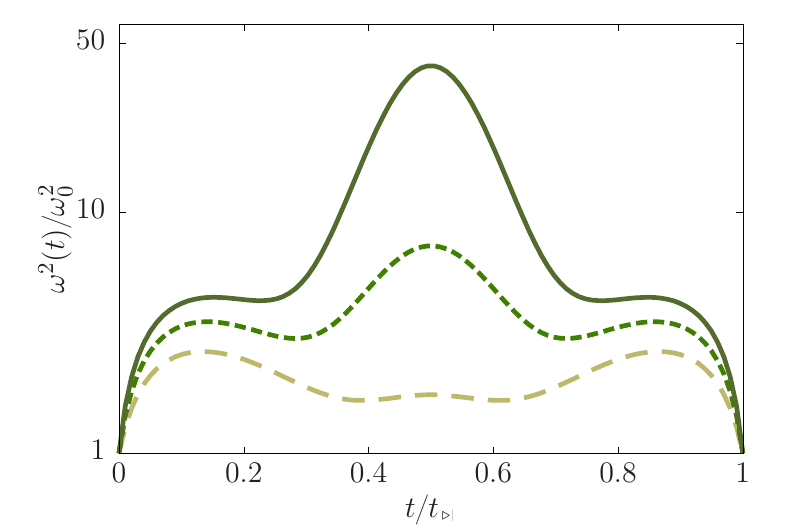}\\
    \caption{Trapping strength
    $\omega^2(t)/\omega_0^2$ as a function of $t/t_{\smalltriangleright\scalebox{.5}{|}}$ for
$\tau(t_{\smalltriangleright\scalebox{.5}{|}})/t_{\smalltriangleright\scalebox{.5}{|}}$=1.77 ($|A|=20$, long-dashed line), 2.80 ($|A|=30$, short-dashed line), 5.56 ($|A|=40$, solid line).
}
\label{fig:tF}
\end{figure}

In Fig. \ref{fig:tF}, we show the behaviour of
$\omega^2(t)/\omega_0^2$ as a function of
$t/t_{\smalltriangleright\scalebox{.5}{|}}$ for different values of
$\tau_{\smalltriangleright\scalebox{.5}{|}}/t_{\smalltriangleright\scalebox{.5}{|}}$
that are determined by the value of $|A|$. We observe that longer time
leaps require larger variations in trapping strength, and
consequently, in density. This must be considered in experimental
realizations to avoid the gas leaving the strongly interacting regime
that is assured for $g m/(\hbar^2 n(x=0,t)) \gg 1$ for all time.

\begin{figure}
    \centering
    \includegraphics[width=1.\linewidth]{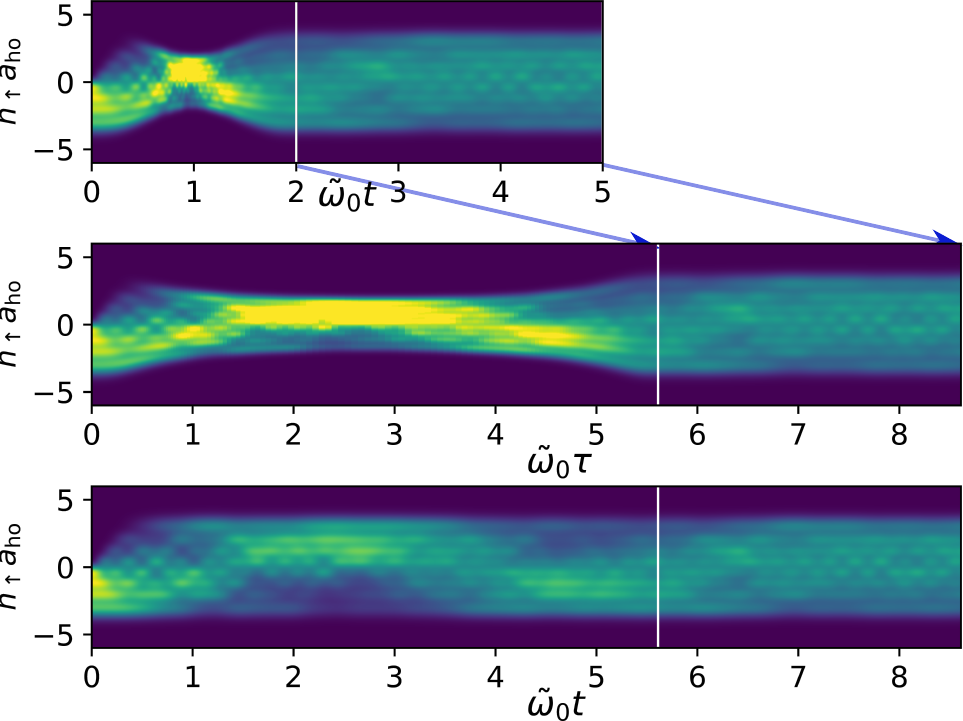}\\
    \caption{Time evolution of the spin density $n_\uparrow(t)$ for an SU(2) fermionic balanced mixture ($N=8$), the initial state being a domain wall. 
    Top panel: Time evolution by applying the 
    shortcut-to-dynamics protocol, as a function of $\tilde\omega_0t$, with $\tilde \omega_0=\hbar\omega_0^2a_{ho}/g$.
    Middle panel: the same as the top panel but a function of $\tilde\omega_0\tau(t)$.
    Bottom panel: Time evolution without applying the  
    shortcut-to-dynamics protocol, as a function of $\tilde\omega_0t$.
    The white vertical lines indicate $t_{\smalltriangleright\scalebox{.5}{|}}$ in the top panel and
    $\tau(t_{\smalltriangleright\scalebox{.5}{|}})$ in the middle and bottom panels.
    The first blue arrow shows the correspondence between $t_{\smalltriangleright\scalebox{.5}{|}}$ and  $\tau_{\smalltriangleright\scalebox{.5}{|}}$,
    while the second one shows the correspondence between the final times $t_F$ and $\tau(t_F)$.
    The spin-down density $n_\downarrow(t)$ (not shown in the figure) is the mirror reflection of $n_\uparrow(t)$ with respect to the trap center for the particular spin excitation considered in this example.
    \label{fig:fast-forward-dens}}
\end{figure}

\begin{figure}
    \centering
    \includegraphics[width=0.8\linewidth]{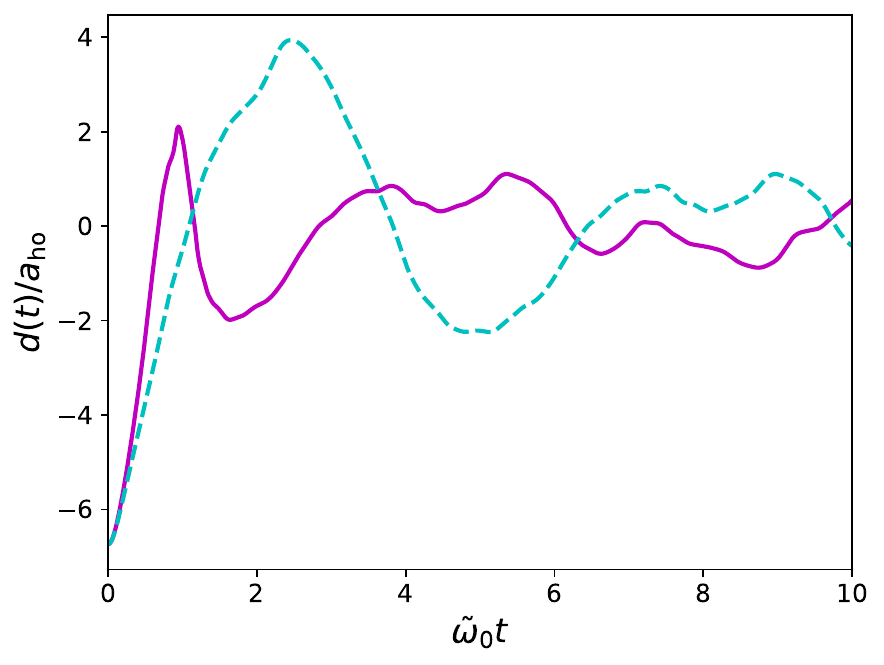}\\
     \includegraphics[width=0.8\linewidth]{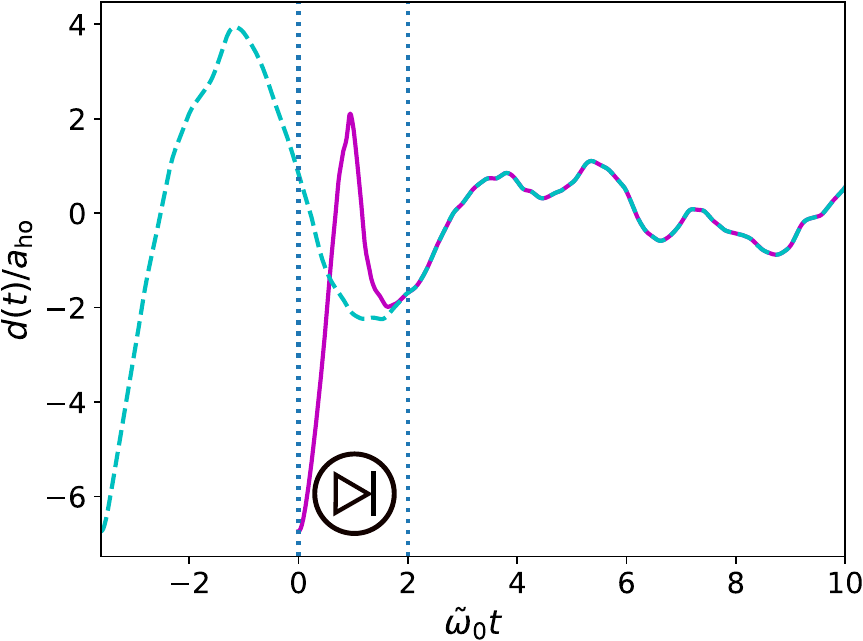}\\
     \caption{Top panel: The magnetization center of mass $d(t)$ as a
       function of $\tilde\omega_0t$, with
       $\tilde \omega_0=\hbar\omega_0^2a_{ho}/g$ for an SU(2) fermionic
       balanced mixture ($N=8$), the initial state being a domain
       wall. We study its time-evolution without exciting the total
       density (cyan dashed curve) and under the action of the
       shortcut-to-dynamics protocol (magenta solid curve).  Bottom
       panel: same as the top panel, but with the cyan curve plotted
       as a function of
       $t-(\tau_{\smalltriangleright\scalebox{.5}{|}}-t_{\smalltriangleright\scalebox{.5}{|}})$
       with
       $\tau_{\smalltriangleright\scalebox{.5}{|}}-t_{\smalltriangleright\scalebox{.5}{|}}$
       being the gained time with the shortcut-to-dynamics
       protocol. The fast-forward time region has been delimited by
       the two vertical dashed lines.}
    \label{fig:fast-forward}
\end{figure}

We have implemented the shortcut-to-dynamics protocol outlined above
for the case $|A|=30$, which corresponds to a maximum change of the
trapping potential of a factor $\sim 3$ and to a maximum variation of
the density at the center of the trap of the $\sim 30\%$ at
$t\simeq t_{\smalltriangleright\scalebox{.5}{|}}/2$.  This choice
fixes
$\tau_{\smalltriangleright\scalebox{.5}{|}}=2.80
t_{\smalltriangleright\scalebox{.5}{|}}$. We have studied the time
evolution of the spin densities (see Fig. \ref{fig:fast-forward-dens}
for $n_\uparrow(t)$) and of the center of mass of the magnetization
$d(t)$ (see Fig. \ref{fig:fast-forward}) for the case of 4+4 SU(2)
fermions initially prepared in a domain wall configuration at $x=0$,
where the spins up and down are located in the two opposite sides of
the trap.  In both Figs. \ref{fig:fast-forward-dens} and
\ref{fig:fast-forward}, we compare the dynamics with and without using
our shortcut-to-dynamics protocol, and we show that the dynamics at a
time $t_{\smalltriangleright\scalebox{.5}{|}}+t'$ in the first case
coincides with that at a time
$\tau_{\smalltriangleright\scalebox{.5}{|}}+t'$ in the second case, as
expected.  It is important to remark that during the
  application of the shortcut-to-dynamics protocol
  ($0<t<t_{\smalltriangleright\scalebox{.5}{|}}$) not only there is a
  fast-forwarding of the dynamics, but also (see central panel of
  Fig. \ref{fig:fast-forward-dens}) a squeezing of the density
  [Eq. (\ref{spindensscaled})]. Thus, if one would like to reconstruct
  the real-time dynamics from the contracted dynamics, both for the
  spin density and the center-of-mass of the magnetization
  [Eq. (\ref{ddt}) and Fig. \ref{fig:fast-forward}], the compression
  by the scaling factor $b(t)$ has to be taken into account in
  addition to the rescaled time $\tau(t)$.

We note that the same shortcut-to-dynamics protocol can be repeated
several times in the same experiment, so that one can have access to
the spin long-time dynamics with a sequence of time leaps.  Let us
underline once more that the same protocol can be applied to any
possible spin excitation in any mixture described by the Hamiltonian
(\ref{ham}) in the limit of strong repulsive interactions, namely in
situations where the total density behaves as that of a system of
non-interacting fermions, and the spin-dynamics is driven by the total
density.

{\it Conclusions -} We have shown that it is possible to jump forward
in the spin dynamics in a strongly repulsive quantum mixture by
rapidly compressing and then decompressing the total density.  The
time $t_{\smalltriangleright\scalebox{.5}{|}}$ of the
compression-decompression cycle is the time needed for the
fast-forward process to advance the spin system to a later time
$\tau_{\smalltriangleright\scalebox{.5}{|}}$ of its natural evolution,
with
$\tau_{\smalltriangleright\scalebox{.5}{|}}\gg
t_{\smalltriangleright\scalebox{.5}{|}}$.  The many-body wavefunction
is identical to the one at time
$\tau_{\smalltriangleright\scalebox{.5}{|}}$ except for a global phase
factor that does not affect any physical observable.  Our protocol is
analytical for the case of a system that is trapped in a harmonic
confinement, but one could look for some numerical protocols for the
case of a strongly interacting mixture in the presence of disorder.
This would pave the way to an easier access, both numerically and
experimentally, to the long-time many-body dynamics of these quantum
systems.

As a final remark, one could wonder if a similar fast-backward
(previous chapter) protocol would be possible.  This would imply a
time $\tau_{\smalltriangleright\scalebox{.5}{|}}$ that is negative,
but it wouldn't be possible by keeping the interaction strength
constant.  Indeed, even if $b(t)$ were negative, the positivity of the
density dictates that $d\tau=dt/|b|^3$. $d\tau$ being always positive,
it is not possible to bring the spin system to its past. However, a
possible strategy could involve sweeping from strongly repulsive to
strongly attractive interaction using a confinement induced resonance
\cite{Astrakarchik2005,Haller2009,Guan2010}. This would change the
sign of the hopping amplitudes $J_i$, and thus the orientation of the
time axis. From an experimental point of view, the difficulty would
reside in the metastability of the gas in the attractive regime, but
this fast-backward strategy deserves to be explored in future works.

\section*{Acknowledgments}
The authors acknowledge the financial support from the CNRS
International Research Project COQSYS, from the Scientific Research
Fund of Piri Reis University under Project Number BAP-2025-02-06,
and from the ANR-21-CE47-0009 Quantum-SOPHA project.  We
acknowledge Fr\'ed\'eric H\'ebert, Christian Miniatura and
Quentin Glorieux for fruitful discussions.

%\bibliography{biblio}

\end{document}